\documentclass{article}

\usepackage{arxiv}

\usepackage[utf8]{inputenc} 
\usepackage[T1]{fontenc}    

\usepackage{url}            
\usepackage{booktabs}       
\usepackage{amsfonts}       
\usepackage{amsmath}
\usepackage{nicefrac}       
\usepackage{microtype}      
\usepackage[title]{appendix}
\usepackage{graphicx}
\usepackage[bf]{caption}
\usepackage{float}
\usepackage{microtype}

\usepackage{doi}
\usepackage{etoolbox}
\usepackage[natbibapa]{apacite}
\AtBeginDocument{}
\renewcommand{\APACrefnote}[1]{}

\newtoggle{bibdoi}
\newtoggle{biburl}
\makeatletter
\newsavebox{\bib@url}
\newsavebox{\bib@doi}

\undef{\APACrefURL}
\undef{\endAPACrefURL}
\undef{\APACrefDOI}
\undef{\endAPACrefDOI}

\newenvironment{APACrefURL}
  {\global\toggletrue{biburl}\lrbox\bib@url}
  {\endlrbox}

\newenvironment{APACrefDOI}
  {\global\toggletrue{bibdoi}\lrbox\bib@doi}
  {\endlrbox}

\newcommand{\printinfo}{
  \iftoggle{bibdoi}{\usebox{\bib@doi}}{\usebox{\bib@url}}
  \togglefalse{bibdoi}
}

\AtBeginEnvironment{thebibliography}{
\pretocmd{\PrintBackRefs}{%
  \iftoggle{bibdoi}
    {\iftoggle{biburl}{\unskip\unskip}{}\usebox{\bib@doi}}
    {\iftoggle{biburl}{Retrieved from \usebox{\bib@url}}}{}
  \togglefalse{bibdoi}\togglefalse{biburl}%
}{}{}}
\makeatother

\title{Dissecting Resilience Triangle: Unravelling Resilience Curve Archetypes and Properties in Human Systems Facing Weather Hazards}

\date{} 					

\renewcommand{\headeright}{}
\renewcommand{\undertitle}{}
\renewcommand{\shorttitle}{}

\hypersetup{
pdftitle={Quantitative Measures for Integrating Resilience into Transportation Planning Practice: Study in Texas},
pdfsubject={},
}

\begin{document}
\maketitle

\begin{center}
{\Large
Chia-Wei Hsu\textsuperscript{a,*}, 
Ali Mostafavi\textsuperscript{a}
\par}

\bigskip
\textsuperscript{a} Urban Resilience.AI Lab, Zachry Department of Civil and Environmental Engineering,\\ Texas A\&M University, 199 Spence St., College Station, TX 77843\\
\vspace{6pt}
\textsuperscript{*} correseponding author, email: chawei0207@tamu.edu
\\
\end{center}
\bigskip
\begin{abstract}
Resilience curves have been the primary approach for conceptualizing and representing the resilience behavior of communities during hazard events; however, the use of resilience curves has remained as a mere conceptual and visual tool with limited data-driven characterization and empirical grounding. Empirical characterizations of resilience curves provide essential insights regarding the manner in which differently impacted systems of communities absorb perturbations and recover from disruptions. To address this gap, this study examines human mobility resilience patterns following multiple weather-related hazard events in the United States by analyzing more than 2000 empirical resilience curves constructed from high-resolution location-based mobility data. These empirical resilience curves are then classified using k-means clustering based on various features (e.g., residual performance, disruptive duration, and recovery duration) into archetypes. Three main archetypes of human mobility resilience are identified: Type I, with rapid recovery after mild impact; Type II, exhibiting bimodal recovery after moderate impact; and Type III, showing slower recovery after severe impact. The results also reveal critical thresholds, such as the bimodal recovery breakpoint at a 20\% impact extent (i.e., function loss), at which the recovery rate decreases, and the critical functional threshold at a 60\% impact extent, above which recovery rate would be rather slow. The results show that a critical functional recovery rate of 2.5\% per day is necessary to follow the bimodal resilience archetype when impact extent exceeds more than 20\%. These findings provide novel and important insights into different resilience curve archetypes and their fundamental properties. Departing from using resilience curves as a mere concept and visual tool, the data-driven specification of resilience curve archetypes and their properties improve our understanding of the resilience patterns of human systems of communities and enable researchers and practitioners to better anticipate and analyze ways communities bounce back in the aftermath of disruptive hazard events. 

\end{abstract}



\section{Introduction}
\label{sec:Introduction}
The characterization of resilience in human systems is of primal importance when evaluating their performance during and after disasters \citep{alexander_resilience_2013,gunderson_ecological_2010,hsu_human_2022,roy_quantifying_2019,wang_patterns_2016}. Studies conducted in recent years, have focused on characterizing resilience curves \citep{kammouh_resilience_2019,gama_dessavre_multidimensional_2016,zobel_characterizing_2014} which graphically represent the trajectory of a community's functionality or performance from the onset of a disaster to the eventual recovery \citep{bruneau_framework_2003,hosseini_review_2016,manyena_concept_2006,panteli_metrics_2017,tierney_conceptualizing_2007}. The resilience curves are the primary approach in understanding and anticipating a community's response to perturbations induced by disasters. However, the use of resilience curves has remained as a mere conceptual and visual tool with limited data-driven characterization and empirical grounding. Empirical characterizations of resilience curves provide essential insights regarding ways different systems of communities absorb perturbations and recover from disruptions \citep{bostick_resilience_2018,ganguly_critical_2018,li_joint_2019}. \\
Resilience curves have also been primarily used in characterizing the vulnerability and recovery of infrastructure systems of communities during hazard events. Limited studies have examined the characteristics of resilience curves in human systems of communities. In recent years, a number of studies have examined fluctuations in human mobility patterns during hazard events as a way to capture both the loss of functionality and subsequent recovery of human systems in communities. \citep{chan_measuring_2016,hong_measuring_2021,kammouh_resilience_2019,platt_measuring_2016,rus_resilience_2018,zhang_resilience-based_2016}. Examining resilience curves associated with human mobility enables the capture of fluctuations in human activities in response to disruptive events, such as floods, wildfires, storms, pandemics, or conflicts \citep{coleman_human_2022,farahmand_anomalous_2022,gao_early_2021,hsu_human_2022,rajput_latent_2022,rajput_latent_2023,tang_resilience_2023}. Human mobility captures the overall functionality of human systems in communities. Under normal circumstances, human mobility would be at an equilibrium state, signifying routine movement patterns. When a disruptive event occurs, human mobility usually decreases as people seek shelter or infrastructure is disrupted, reflected by a dip in the resilience curve. Post-disaster, human mobility gradually recovers as people start to adapt, recover, and resume their routines \citep{nicholson_flow-based_2016}. The fluctuations in human mobility patterns can be captured to construct the resilience curve of human systems of communities. Despite the growing number of studies examining resilience of human systems using human mobility patterns, limited attention has been paid to the resilience curve archetypes and their fundamental properties using empirical data. The specification of empirical resilience curve archetypes and their properties is essential to improve our understanding of the resilience patterns of human systems of communities and to enable researchers and practitioners to better anticipate and analyze ways communities bounce back in the aftermath of disruptive hazard events \citep{chang_measuring_2004,gao_resilience-oriented_2017}.\\
Recognizing this important knowledge gap, the primary objective of this study is to examine the presence of universal archetypes in human mobility resilience curves and delineate their fundamental properties. Specifically, we seek to address two specific research questions: (1) What are the primary archetypes of the human system resilience curves? and (2) What fundamental characteristics explain the behavior of different resilience curve archetypes? To assess the extent of functionality loss, we measure the degree of human mobility change by computing the number of trips going in and out of a given area. For our analysis, we utilize high-resolution location-based mobility data related to multiple extreme weather events in the United States. In total, we constructed more than 2000 empirical resilience curves representing different regions and hazard events. Accordingly, we examined and computed the main features of each resilience curve, and subsequently classified them based on their main features into a set of universal archetypes. The following sections discuss the study data and methods.

\section{Data Description}
\label{sec:Data Description}
This study collected and analyzed data from the following major hazard events in the United States: Hurricane Ida, Hurricane Harvey, Hurricane Laura, and Winter Storm Uri. Hurricane Ida, a Category 4 storm, struck in August 2021, causing its most profound devastation in Louisiana.  Hurricane Ida later moved on to cause significant flooding and damage in the northeastern U.S., particularly impacting New York and New Jersey. Hurricane Harvey, also a Category 4 storm, made landfall near Rockport, Texas, in August 2017, then unleashed catastrophic flooding on the Houston metropolitan region. The impact was so severe that it ranks among the most damaging natural disasters in U.S. history. Category 4 storm Hurricane Laura in August 2020 majorly affected the Gulf Coast of the U.S., impacting parts of Louisiana and Texas and leaving a trail of destruction in its wake. In 2021, Winter Storm Uri swept across multiple states, with Texas being the most affected. Bringing low temperatures, snow, and ice, Uri led to extensive power outages, immense property damage, and several tragic fatalities. The storm's severity was such that Texas's power grid was overwhelmed, resulting in widespread blackouts across the state. These four events were chosen due to their significant disruptions across the impacted regions. The data collection timeframes for each event were established according to the event's start date. Specifically, data collection commenced 9 days before the initiation of extreme weather events and spanned a total 35 days for such occurrences. In the case of the winter storm event, the data collection timeframe was limited to 24 days due to data availability constraints. The baseline periods—normal (steady-state) periods without perturbations from natural phenomena disasters—are set in the first week of the data collection timeframes. The mobility flow during this period is viewed as baseline performance. Table 1 summarizes the data collection details for each event. 

\begin{table}[!ht]
    \caption{Summary of the weather events selected for this study}
    \centering
    \begin{tabular}{llllll}
    \hline
        Event & Type & Affected areas & \begin{tabular}[c]{@{}l@{}}Number of\\ census tracts\end{tabular} & Event start date & \begin{tabular}[c]{@{}l@{}}Data collection\\ time frame\end{tabular}\\ 
        \hline
        Ida & Hurricane & Louisiana & 402 & 2021/8/28 & 2021/8/19–2021/9/23  \\ 
        Harvey & Hurricane & Texas & 786 & 2017/8/24 & 2017/8/15–2017/9/30  \\ 
        Laura & Hurricane & Louisiana & 402 & 2020/8/27 & 2020/8/18–2020/9/22  \\ 
        Uri & Winter Storm & Texas & 786 & 2021/2/13 & 2021/2/4–2021/2/28  \\ 
        \hline
    \end{tabular}
\end{table}

To construct the empirical resilience curves, we used a location-based dataset obtained from Spectus, a company that collects vast amounts of anonymous location information from approximately 70 million mobile devices in the United States through a privacy-compliant framework. This data is gathered when users voluntarily opt in to location services provided by partner apps. Spectus captures data from nearly 20\% of the U.S. population, representing around one in four smartphone users. The location-based data from Spectus has proven valuable in previous research \citep{hsu_human_2022,hsu_human_2022-4} to be representative in terms of capturing human mobility and travel mode detection due to its high spatiotemporal resolution. To safeguard user privacy, Spectus de-identifies the collected data and applies additional privacy measures, including obscuring home locations at the census-block group level and removing sensitive points of interest. The device-level data encompasses anonymized individual location information, ID, location coordinates, and timestamps. To ensure privacy preservation, Spectus offers access to its data, tools, and location-based datasets from other providers through a data cleanroom environment. The locations visited by each device are determined by identifying the census tract polygons each device resides in. The trajectory of a device's movement is then established based on the precedence relationship of their visit times. Daily trip counts between census tracts are aggregated at a census-tract level. For further analysis, a human mobility network is created using the daily trip counts among each census-tract pair. In this study, the focus lies on examining the fluctuation of total mobility flow in each census tract. Inflows and outflows are aggregated, as their separation within the period of interest does not yield additional information for characterizing resilience curves. This is because inflow and outflow are proportional to the total flow and share the same flow pattern.

\section{Methodology}
\label{sec:Methodology}
Our analysis framework consists of the following components: (1) constructing the resilience curves; (2) extracting the key features of resilience curves; (3) finding the universal resilience curve archetypes and (4) specifying the main properties of the archetypes. Figure 1 depicts the overview of the analysis framework.\\

\begin{figure}[!ht]
	\centering
	\includegraphics[width=0.8\textwidth]{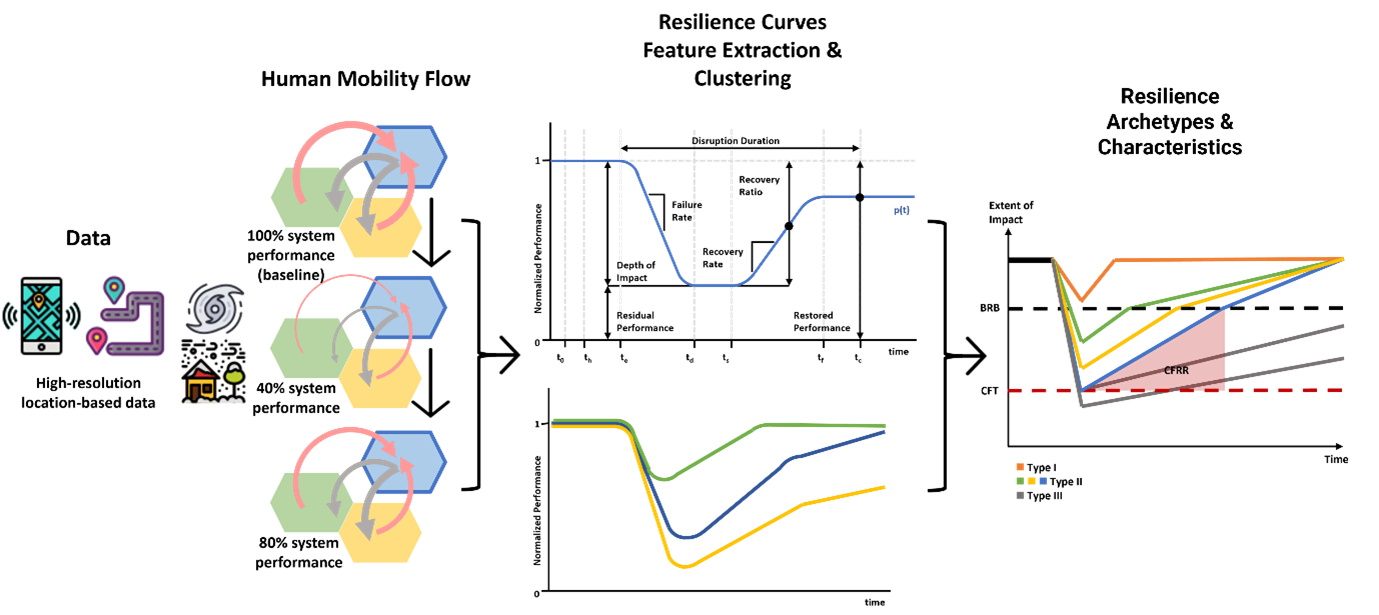}
	\caption{Overview of analysis framework: Location-based data are analyzed to estimate mobility flow as a proxy for the functionality of human systems. Key resilience features, including the critical transition time points, are extracted from the human mobility resilience curves for each census tract. Based on their features, resilience curves were clustered to specify universal archetypes. The archetypes for human mobility resilience curves are examined to delineate their fundamental properties.}
	\label{fig:fig1}
\end{figure}

Our initial step focused on constructing empirical resilience curves associated with human mobility of census-tract populations across the four hazard events examined in the study. These empirical resilience curves are constructed by comparing the number of trips each day within the 35 days data-collection timeframe with the baseline (steady-state) number of trips for each census tract. The baseline number of trips serves as the baseline functionality of each census tract. For example, if the number of trips observed on a certain day during Hurricane Ida is 70\% of the baseline value, then the remaining system functionality is 70\%; the impact extent is 30\%. Figure 2 shows a conceptual resilience curve, critical temporal points that signal transitions, and resilience features characterizing the curve. Each resilience curve is composed of critical temporal points: th (exposure to hazard): the time when the system first experiences the disruptive event; te (initial system disruption): the time when the system experiences the maximum disruptive effects; td (end of cascading failures): the time point when disruption starts to diminish; ts (beginning of system recovery): the onset of the system's recovery period; tf (completion of system recovery): the time point when the system is considered to have fully recovered and tc (maximum recovery time): the complete end of the event, marking a return to normal conditions. These points encapsulate the timeline of the event and the corresponding system performance.\\
The subsequent step involves extracting key features from the resilience curves. The selection of key features is based on a review of the literature related to the characteristics of resilience curves (Table 2) \citep{hillebrand_decomposing_2018,poulin_infrastructure_2021}. The key features collected in the study can be grouped into multiple categories to mitigate noise and enhance the curves' discernibility, we employed a Savitzky-Golay filter \citep{press_savitzkygolay_1990,savitzky_smoothing_1964,simonoff_smoothing_2012} after comparing the performance of multiple common smoothing techniques, such as rolling average and interpolators. However, curve smoothing should be applied judiciously to avoid potential over-smoothing, which could obscure important details. To reliably record the system resilience performance, we use the smoothed curve only for computing features related to integral and rate features and computed the other features from the original resilience curves. Upon completion of this step, the resilience curve for each census tract is represented by a vector of multiple key resilience features.\\

\begin{figure}[!ht]
	\centering
	\includegraphics[width=0.5\textwidth]{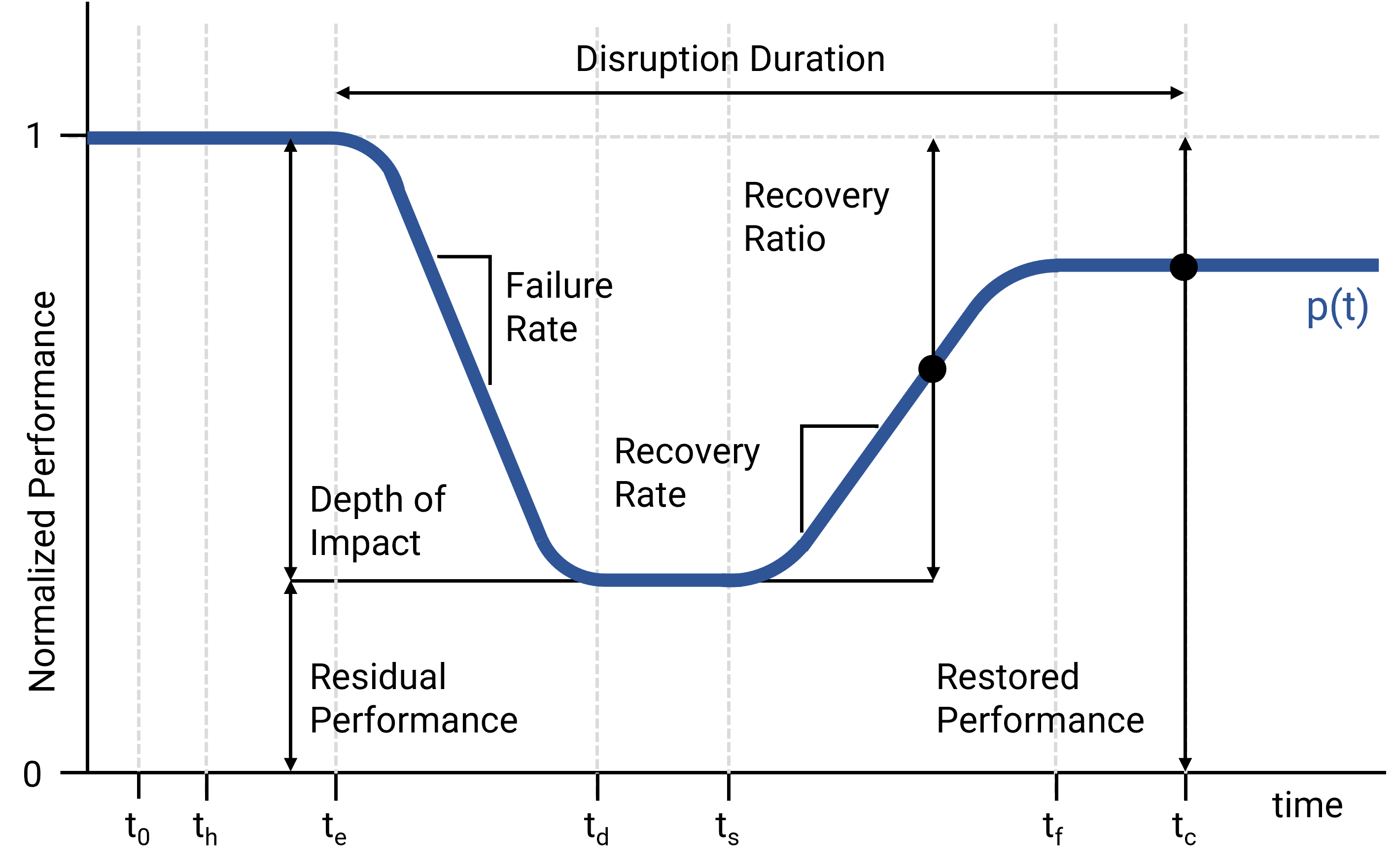}
	\caption{Illustrative features for a conceptual resilience curve. Metrics may be derived with respect to performance p(t). In this illustration, the system does not fully recover within the control interval, so disruptive duration may be undefined. The area above the resilience curve represents cumulative impact, while the area under the resilience curve represents cumulative performance.}
	\label{fig:fig2}
\end{figure}

\begin{table}[!ht]
\caption{Key resilience features extracted from the resilience curve regarding human mobility recovery for clustering.}
\centering
\begin{tabular}{p{2cm}p{3.5cm}p{5cm}p{5cm}}
\hline
\textbf{Types} & \textbf{Metrics} & \textbf{Formula} & \textbf{Definition} \\
\hline
Magnitude & Residual performance & \(p(t_d)\) & System performance following the disruption, generally after cascading failures. \\
& Depth of impact & \(1-p(t_d)\) & Complement of residual performance. \\
& Restored performance & 
\(\frac{p(t_f)}{p(t_e)}\) \(\frac{p(t_f) - p(t_d)}{p(t_e) - p(t_d)}\) & System's performance after recovery efforts are complete. \\

Duration & Disruptive duration & \(t_f - t_e\) & Entire period of degraded performance. \\
& Recovery duration & \(t_f - t_d\) & Period of the recovery phase, starting from the lowest performance. \\

Integral & Cumulative impact & \(\int{1 - p(t)}dt\) & Integrated difference between performance and its reference. \\
& Cumulative performance & \(\int p(t)dt\) & Complement of cumulative impact. \\ \\

Rate & Failure rate & \(\frac{p(t_d) - p(t_e)}{t_d - t_e}\) & Resilience and adaptive capability at failure phases. \\
& Recovery rate & \(\frac{p(t_f) - p(t_s)}{t_f - t_s}\) & Restorative capability at recovery phases. \\

Stability & Temporal stability & \begin{tabular}[c]{@{}p{5cm}@{}} $d=1/std($\text{residual}$_b)$ \\ \(\ln\left(\frac{\text{recovery\ rate}}{\text{average\ recovery\ rate}}\right) = i + b \cdot t\) \end{tabular} & Performance fluctuations around the trend. No benchmark, a larger \(d\) corresponds to lower fluctuations. \\
\hline
\end{tabular}%
\end{table}

Next, we perform clustering algorithms to classify resilience curves based on the key features to examine if these clusters would represent different universal archetypes. For testing different clustering algorithms, we used the elbow method and silhouette scores. Based on these metrics, we chose k-means clustering, which is a widely used technique due to its efficiency and simplicity. The k-means \citep{ikotun_k-means_2023,jin_k-means_2010,han_estimating_2009,kanungo_efficient_2002,likas_global_2003,steinley_k-means_2006} algorithm partitions the data into k distinct, non-overlapping subsets (or clusters), with each data point belonging to the cluster with the closest mean. After clustering, we proceed to computing an average resilience curve for each cluster. This step provides us with a representative curve that encapsulates the typical behavior for each cluster. These representative curves serve as resilience curve archetypes.\\
In the last step, we aim to find out the fundamental properties of the resilience curve archetypes. We apply multivariate adaptive regression splines (MARS) \citep{balshi_assessing_2009, friedman_multivariate_1991, kisi_application_2016,miao_road_2013} for piecewise linear regression to simplify the representative resilience curves of each cluster and to reveal their common structures. MARS is a form of nonparametric regression technique that identifies the main turning points and slopes (system performance change rates). We use simplified curves to specify their main properties of resilience curve archetypes.

\section{Results}
\label{sec:Results}
We implemented the method described in the previous section on the data collected from Hurricane Ida, Hurricane Harvey, Hurricane Laura and winter storm Uri. Evaluation metrics for k-means clustering such as the elbow method and silhouette score suggest that the optimum number of clusters for the events is six. Among the six clusters, one or two clusters can be viewed as outliers which should be removed for the interpretation of results. After the clustering and cleaning, we performed MARS to identify the critical turning points as the slope of line segments between these points. Figures 3 through 6 show the average resilience curve for each cluster, critical turning point and slopes of the line segments between the turning points. The x-axes of the plots represent the dates within our data collection timeframe while y-axes represent the system functional performance.\\
The shapes of resilience curves are all triangular instead of trapezoidal, due to the more adaptive nature of human systems compared to infrastructure systems. Human mobility starts to recover immediately after the shock while infrastructure systems may experience a sustained period of impact before their functionality starts recovering. The areas with larger impact are either the ones with communities with larger populations or coastal areas (specifically for hurricane events), which is intuitive.\\
Figure 3 shows the clustering results for Hurricane Ida in which the extent of impact varies, ranging from 20\% to 70\%. We can observe that except for cluster 5, all the other clusters fully recover at some point within our data collection timeframe. Based on this result, if the impact extent exceeds 70\%, then a region may take much longer to fully recover, which is not captured due to our data collection timeline. Except for cluster 3, we observe a significant slow-down in the recovery rate after passing a certain level of system performance in all the other clusters, 20\% in this case.

\begin{figure}[!ht]
	\centering
	\includegraphics[width=0.7\textwidth]{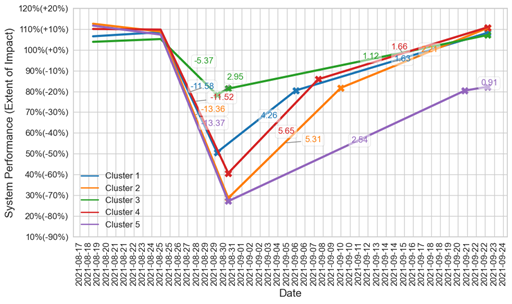}
	\caption{Human mobility resilience curve clustering for Hurricane Ida. Each curve represents the MARS regression on the average resilience curve of each cluster; the numbers represent failure rates and recovery rates. The extent of impact ranges from 20\% to 70\%. The turning point of bimodal recovery rate can be seen in cluster 1, 2 and 4 when the system recovers to between 80\% to 90\% of performance. Cluster 3 represents areas with relatively low impact and fast recovery. Cluster 5 represents areas with large impact and slow recovery.}
	\label{fig:fig3}
\end{figure}

Figure 4 shows the clustering results for Hurricane Harvey; the extent of impact ranges from 55\% to 70\%. We can observe that except for Cluster 1, all the other regions fully recover at some point within our data collection timeframe. For all the clusters, we see a significant decrease in the recovery rate after achieving a certain level of system functional performance, 20\% in this case. 

\begin{figure}[!ht]
	\centering
	\includegraphics[width=0.7\textwidth]{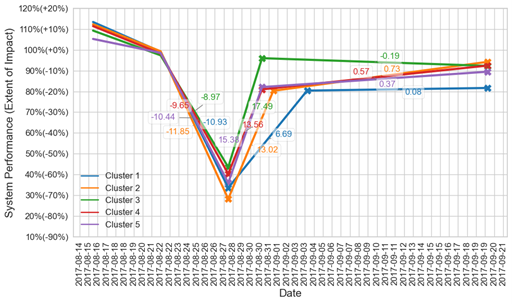}
	\caption{Human mobility recovery clustering for Hurricane Harvey. Each curve represents the MARS regression on the average resilience curve of each cluster; the numbers represent failure rates and recovery rates. The extent of impact ranges from 55\% to 70\%. The turning point of bimodal recovery rate can be seen in clusters 1, 2, 4, and 5 when the system recovers to between 80\% to 90\% of performance. Cluster 3 represents areas with relatively low impact and fast recovery.}
	\label{fig:fig4}
\end{figure}

Figure 5 shows the clustering result for Hurricane Laura; the extent of impact ranges from 50\% to 85\%. We note that all regions, with the exception of cluster 3, show complete recovery within our data collection period. Additionally, after reaching a specific threshold of system functionality, 20\% in this instance, there is a noticeable decline in the recovery speed across the other clusters.

\begin{figure}[!ht]
	\centering
	\includegraphics[width=0.7\textwidth]{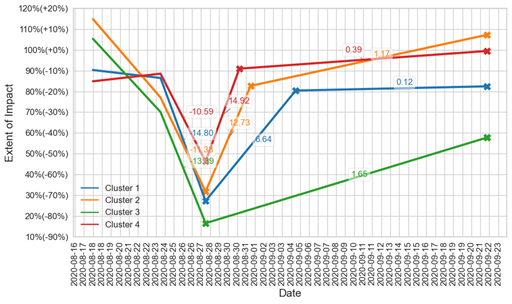}
	\caption{Human mobility recovery clustering for Hurricane Laura. Each curve represents the MARS regression on the average resilience curve of each cluster; the numbers represent failure rates and recovery rates. The extent of impact ranges from 50\% to 85\%. The turning point of bimodal recovery rate can be seen in clusters 1, 2, and 4, when the system recovers to between 80\% to 90\% of performance. Cluster 3 represents areas with large impact and slow recovery.}
	\label{fig:fig5}
\end{figure}

Figure 6 shows the clustering result for Winter Storm Uri; the extent of impact ranges from 40\% to 60\%. We notice that every region achieves full recovery within the duration of our data collection; however, once they reach a system functional performance threshold of 20\%, there is a marked slowdown in their pace. 

\begin{figure}[!ht]
	\centering
	\includegraphics[width=0.7\textwidth]{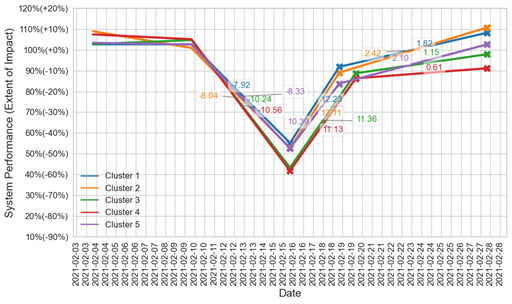}
	\caption{Human mobility recovery clustering for Winter Storm Uri. Each curve represents the MARS regression on the average resilience curve of each cluster; the numbers represent failure rates and recovery rates. The extent of impact ranges from 40\% to 60\%. The turning point of bimodal recovery rate can be seen when the system recovers to between 80\% to 90\% of performance.}
	\label{fig:fig6}
\end{figure}

The results for the selected events revealed multiple important insights regarding fundamental properties of resilience curve archetypes of human mobility. Figure 7 shows the conceptual resilience curve resilience archetypes found in our empirical study. The three archetypes are as follows: Type I: This archetype comprises areas that experienced the least impact and exhibited relatively rapid recovery. Human mobility in these regions resumed quickly, showcasing efficient recovery processes. Type II: Areas with resilience curves of this archetype encountered moderate levels of impact. Notably, we observed a bimodal recovery rate, where the system's recovery speed significantly slowed down upon reaching a certain functional performance level. This bimodal phase transition was distinct from the initial recovery and had a noticeable impact on the overall recovery process. Type III: Representing the most affected areas, this group demonstrated considerably slower recovery rates compared to the other archetypes. It took significantly longer for these regions to return to full system performance. 
Further investigation into these three archetypes led to identification of significant critical thresholds and distinguishing properties of the archetypes. 
Bimodal recovery breakpoint (BRB): The BRB is the point at which the recovery rate changes. The BRB was specified at the impact extent of 20\%. If the impact extent is less than the BRB, human mobility recovers swiftly after hazard perturbations. If the impact extent is greater than the BRB, however, human mobility initially recovers to the BRB level with a faster recovery rate; after the BRB is achieved, the recovery rate slows down.
Critical functional threshold (CFT): The CFT is the impact extent beyond which the recovery of the system would proceed at a slow rate causing long recovery durations. The CFT for human mobility was identified with the impact extent of 60\%. An impact extent greater than this threshold would lead to significantly slower recovery speed. The initial recovery rate right after impact (RR1) needs to follow the critical functional recovery rate (CFRR) which is 2.5\% per day. The extent of impact played a significant role in determining the recovery behavior, particularly when it is greater than the CFT. Table 3 summarizes properties of different resilience curve archetypes related to human mobility and their fundamental properties. In summary, our examination of these distinct archetypes sheds light on resilience behavior of human mobility in hazard events. The identification of the BRB and CFT provides valuable insights into the recovery patterns of different areas, facilitating better anticipation and evaluations of ways different areas recover from hazard-induced perturbations.\\

\begin{figure}[!ht]
	\centering
	\includegraphics[width=0.7\textwidth]{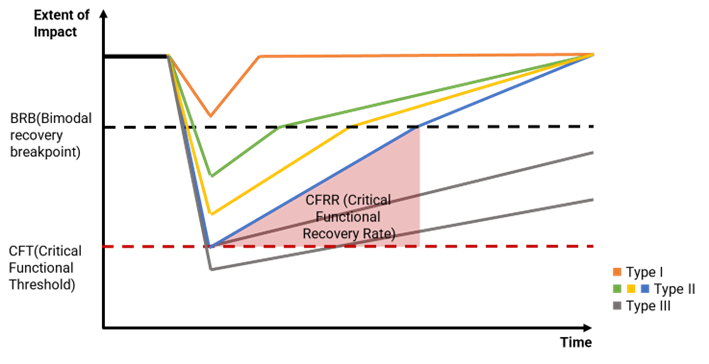}
	\caption{Conceptual representation of the archetypes of human mobility recovery behavior post weather events. Each colored curve represents a cluster found in the analysis which is then further categorized into three archetypes. Critical functional threshold, bimodal recovery breakpoint, and critical functional recovery rate are the governing characteristics that separate the archetypes.}
	\label{fig:fig7}
\end{figure}

\begin{table}[!ht]
\caption{The three archetypes of human mobility recovery behavior following weather events, and the governing characteristics that distinguish archetypes.}
\centering
\begin{tabular}{p{3cm}p{3.5cm}p{3.5cm}p{3.5cm}}
\hline
\textbf{Archetypes} & \textbf{Extent of Impact} & \textbf{Recovery Rate} & \textbf{Description} \\
\hline
Type I & \(I < BRB\) & \(RR_1 > CFRR\) & Least impact \\
& & & Fastest recovery \\
& & & Fully recovered \\

Type II & \(CFT > I > BRB\) & \(RR_1 > CFRR\) & Moderate impact \\
& & & Bimodal recovery rates \\
& & & Fully recovered \\

Type III & \(I > CFT\) & \(RR_1 < CFRR\) & Largest impact \\
& & & Slowest recovery \\
\hline
\multicolumn{4}{p{14cm}}{
\(I\): Extent of impact \newline
\(\text{RR1}\): Initial recovery rate right after impact \newline
\(\text{BRB}\): Bimodal recovery breakpoint \newline
\(\text{CFT}\): Critical functional threshold \newline
\(\text{CFRR}\): Critical functional recovery rate
}
\end{tabular}
\end{table}

\section{Discussion and Concluding Remarks}
\label{sec:Discussion and Concluding Remarks}
The primary objective of this study is to explore the existence of universal archetypes in human mobility resilience curves when communities face weather hazards. Over the past two decades, resilience curves have been widely used to characterize the fluctuations in the functional performance of human and physical infrastructure systems of communities during disruptions; however, the majority existing characterizations of resilience curves of communities do not consider variations in the types of resilience curve patterns a system might exhibit depending on the extent of impact. Also, the majority of existing resilience curve characterizations focus on physical infrastructure; limited studies have examined the characteristics of resilience curves in human systems of communities in an empirical manner. 
Recognizing these important gaps, this study examined more than 2000 empirical resilience curves related to human mobility behaviors across multiple geographic regions and various weather events. The analyses examined datasets collected from Hurricanes Ida, Harvey, and Laura, and Winter Storm Uri in the United States. The results show that the resilience curves can generally be divided into three universal archetypes with low, medium, and high impact extent. The group with low impact exhibited the fastest recovery speed. In the second archetype with moderate impact extent, the recovery follows a bimodal rate: the initial recovery to the bimodal recovery breakpoint proceeds at a faster rate followed by and a slower recovery rate after a breakpoint. This finding suggests that if the impact extent is not severe, human systems of community would strive to recover to the BRB as fast possible. After the system functionality reaches BRB, the recovery rate slows. The existence of the bimodal recovery pattern might be a combination of two different behaviors. In the first stage of recovery, system functional performance quickly bounces back to around 80\% to 90\% of their normal functional performance because a large proportion of their daily activities return to normal. When this level of functional performance is achieved, the human system has a functional performance state, thus the remaining recovery would follow a slower rate. In the third resilience archetype and when the impact exceeds critical functional threshold (i.e., 70\%), the recovery rate would follow a slow rate with a consistent slope. This finding suggests that the critical functionality threshold of 70\% is the point beyond which the human system would struggle to recover. There are differences between the clusters of resilience curves observed from different types of events and different geographical areas; however, the three archetypes revealed in this study show good representativeness covering all possible post-event human mobility recovery behaviors. 
The study findings provide multiple important scientific and practical contributions. First, resilience curves have remained a mere conceptual and visual tool for understanding fluctuations in behaviors of community systems during and after hazard events. The absence of empirical grounding and specific characterization of resilience curves have hindered the ability to properly analyze and understand recovery trajectories of community systems. The findings of this study reveal the presence of universal resilience curve archetypes with specific properties (i.e., bimodal recovery breakpoint, critical functional threshold, and bimodal recovery rates) that enable evaluation of the way community systems behave in the aftermath of hazard-induced perturbations. Second, departing from the majority of the existing studies that focus on characterizing resilience in physical infrastructure, this study examined the resilience of human systems based on fluctuations in human mobility. By leveraging fine-grained location-based data from multiple hazard events, this study evaluated the functional performance of human systems of communities based on fluctuations in mobility flows to reveal universal resilience curve archetypes. 
From a practical perspective, the outcomes of this study provide important insights for emergency managers and city officials. Based on the understanding of which areas are likely to belong to a certain archetype in an extreme weather event, we can anticipate the functional performance and recovery patterns. With the insights about the extent of impact and recovery trajectory for a short period after impact, decision-makers can take proactive actions to restore infrastructure and allocate resources to areas that are expected to follow a slow recovery rate. By assessing the extent of impact and recovery trajectory during the immediate post-event period, decision-makers can gain insights into an area's future performance. These contributions move us closer to a deeper understanding and a greater predictive insight regarding the resilience behaviors of community systems in hazard events. Future studies can build upon the findings of this study to examine empirical resilience curves and their characteristics in other community systems (such as infrastructure systems) to depart from using resilience curves as a mere conceptual and visual tool and reveal data-driven and empirical characteristics of resilience curves in different systems. For example, future studies can examine empirical data from other systems to evaluate whether similar universal resilience curve archetypes exist in other systems and if they exhibit properties such as critical functionality threshold and bimodal recovery rate. Such insights would move the field of community resilience forward with empirical evidence needed to have a deeper and more predictive understanding of the way different community systems behave during hazard-induced perturbations.

\section*{Acknowledgments}
\label{sec:Acknowledgments}
This material is based in part upon work supported by the National Science Foundation under CRISP 2.0 Type 2 No. 1832662 and CAREER 1846069. The authors also would like to acknowledge the data support from Spectus. Any opinions, findings, conclusions, or recommendations expressed in this material are those of the authors and do not necessarily reflect the views of the National Science Foundation or Spectus.

\section*{Author Contributions}
\label{sec:Author Contributions}
All authors critically revised the manuscript, gave final approval for publication, and agree to be held accountable for the work performed therein. C.H. was the lead Ph.D. student researcher and first author, who was responsible for supervising data collection, performing final analysis, and writing the majority of the manuscript. A.M. was the faculty advisor for the project and provided critical feedback on the project development and manuscript.

\section*{Data availability}
\label{sec:Data availability}
All data were collected through a CCPA- and GDPR-compliant framework and utilized for research purposes. The data that support the findings of this study are available from Spectus (https://spectus.ai/product/), but restrictions apply to the availability of these data, which were used under license for the current study. The data can be accessed upon request submitted to the providers (Spectus representative: Brennan Lake; email: blake@spectus.ai). The data was shared under a strict contract through Spectus’ academic collaborative program, in which they provide access to de-identified and privacy-enhanced mobility data for academic research. All researchers processed and analyzed the data under a non-disclosure agreement and were obligated not to share data further or to attempt to re-identify data.

\section*{Code availability}
\label{sec:Code availability}
The code that supports the findings of this study is available from the corresponding author upon request.

\bibliography{template}  






\end{document}